\documentclass[format=sigconf]{acmart}

\usepackage{hyperref}
\usepackage{natbib}
\usepackage{booktabs} 
\usepackage[utf8x]{inputenc} 
\usepackage{soul}

\def\BibTeX{{\rm B\kern-.05em{\sc i\kern-.025em b}\kern-.08emT\kern-.1667em\lower.7ex\hbox{E}\kern-.125emX}}
\copyrightyear{2020}
\acmYear{2020}
\setcopyright{rightsretained}
\acmConference[CUI '20]{2nd Conference on Conversational User Interfaces}{July 22--24, 2020}{Bilbao, Spain}
\acmBooktitle{2nd Conference on Conversational User Interfaces (CUI '20), July 22--24, 2020, Bilbao, Spain}\acmDOI{10.1145/3405755.3406144}
\acmISBN{978-1-4503-7544-3/20/07}

\begin{document}

%
\title{Finally a Case for Collaborative VR?: The Need to Design for Remote Multi-Party Conversations}

\fancyhead{}

%

\author{Anna Bleakley}
\affiliation{University College Dublin}
\email{anna.bleakley@ucdconnect.ie}

\author{Vincent Wade}
\affiliation{Trinity College Dublin, Ireland}
\email{vincent.wade@adaptcentre.ie}

\author{Benjamin R. Cowan}
\affiliation{University College Dublin}
\email{benjamin.cowan@ucd.ie}

%

%
\begin{abstract}
Amid current social distancing measures requiring people to work from home, there has been renewed interest on how to effectively converse and collaborate remotely utilizing currently available technologies. On the surface, VR provides a perfect platform for effective remote communication. It can transfer contextual and environmental cues and facilitate a shared perspective while also allowing people to be virtually co-located. Yet we argue that currently VR is not adequately designed for such a communicative purpose. In this paper, we outline three key barriers to using VR for conversational activity : (1) variability of social immersion, (2) unclear user roles, and (3) the need for effective shared visual reference. Based on this outline, key design topics are discussed through a user experience design perspective for considerations in a future collaborative design framework.      
\end{abstract}

%
%

%


%

%
\maketitle

\section{Introduction}
Amid the surge in remote working conditions due to current social distancing measures, VR could be seen to provide much promise as a tool to support multiparty communication. As early as the 90s VR spaces have been built to facilitate collaboration ~\cite{anderson1995building,benford1995user} encouraging a move towards multi-party virtual communication. More recently VR platforms have facilitated a multitude of conversational activities including conferences, social networking, and teleconferencing. However, the technology is yet to be adopted on a larger scale due to shortcomings in facilitating multi-party collaborative experiences. 

Currently, there exists a shortage of studies that explore how to better design collaborative VR platforms. This is essential in order to identify concepts crucial to activities and conversations that are a part of virtual collaborative experiences. In the literature, current design choices have been examined ~\cite{jonas2019towards,mcveigh2019shaping,jerald2015vr} but focus mainly on single user experiences or an available platform's features. Some research has begun to study how design decisions support multi-party social interaction ~\cite{roth2018beyond, scavarelli2019towards} but are not targeted at creating design principles applicable for collaboration or in supporting conversations that take place in various contexts. Defining design requirements are necessary to ensure that VR designers build effective collaborative experiences that can extend to a multitude of scenarios. In this provocation, we argue that clear insights to inform the design of VR spaces for effective conversation and collaboration are currently lacking. Our paper highlights three key issues needing to be addressed in collaborative VR research with the hopes of laying the groundwork for future design research in this domain.
\section{Key Issues}
\subsection{Variability of social immersion}
Establishing a sense of presence is a common goal for VR applications and has regularly been used as a user experience measure when assessing VR applications. In VR, users do tend to attain a sense of social presence; a sense of \textit{being there} or in groups and \textit{being there together}. Increasing social presence has the potential to promote social cues common in face-to-face interactions such as eye-contact, joint attention as well as non-verbal and verbal cues relevant to a given context enhancing the effectiveness of communicative activities ~\cite{roth2018beyond}. Yet the level of social presence varies significantly depending on the type of VR application used ~\cite{podkosova2018co}. 

The way avatars or environments are augmented also impact co-presence. In a recent experiment, co-presence is shown to be impacted by level detail of avatars as well as fidelity of the background ~\cite{jo2016effects}. This suggests that multiple design variables are likely to contribute to social immersion and  even affect emotional response in users as demonstrated in \cite{jo2017effects}. Even though significant work has concentrated on this concept, we are still unclear what aspects of the environment, social behaviors, and user appearance impact levels of social presence systematically and how these play out in multi-party contexts. 

\subsection{Unclear collaborative roles}
Signals about one’s identity and role influence social interaction. For example the way an avatar or agent is depicted in these virtual environments is known to affect emotional states and empathic behaviour ~\cite{johnson2018assessing}. For multi-party collaboration, roles are dynamic, with speakers taking on different capacities such as speakers, addressees, by-standers, and overhearers during an interaction ~\cite{pejsa2017me}. Depending on the context for multi-party conversation, the recognition and appropriate signalling of these roles in VR is crucial to consider. As clear conversational roles allow for effective multi-party interactions~\cite{pejsa2017me}, the display of effective conversational mechanics by avatars such as attentiveness and gaze behaviour as well as mechanisms to signal roles within a conversation may lead to more effective turn-taking beyond deploying 'floor-control' policies like tradition conferencing systems. What is more, as agents also become communicative actors in VR spaces, taking on previously human roles such as a facilitators or guides, we must also be cognisant of how their role in the conversation is communicated and how their conversational capabilities can be more clearly signalled through design. 

\subsection{The need for an effective shared visual reference}
Shared visual information on the state of the task as well as state of a remote partner are integral for effective remote collaboration when designing virtual environments ~\cite{smith2018communication}. When performing tasks in physical environments, a shared workspace has been shown to assist users in efficiently collaborating when relying on visual information rather than verbally confirming their actions ~\cite{fussell2000coordination}. Various collaborative multi-party tasks require a shared object of interest such as a Kanban board, whiteboard, or poster - that collaborators refer to physically and verbally. In physical collaboration tasks, speech and action have been shown to relate to mutual visual references in the background that coordinate these actions through non-verbal media ~\cite{flor1998side}. This needs to be considered more effectively for VR based collaboration and contextualized for wider use. 

To our knowledge studies on shared artifacts are limited on a general level, and those available are specific to domains such as human-data interaction~\cite{garcia2019collaborative}, scientific communities ~\cite{garcia2019collaborative},or architectural design spaces \cite{nguyen2016applying} to name a few. In these studies much of the emphasis has been on building virtual environments to accommodate specific tasks rather than conversations that take place in reference to shared artifacts. Although some work has examined collaboration on a general level ~\cite{wolff2007review, otto2006review, roberts2004supporting}, they mainly focus on technological factors or are outdated with the release of new commercial VR applications. Experiences are bound to differ depending  on differences in artifacts and settings in question as different contexts and tasks will have specific needs to shape the form of interactions most effective. For instance, a joint programming workflow will require vastly different shared coordination tools and visual reference artifacts in contrast with a joint learning context. The design of visual references therefore need to be understood with consideration to how mutual objects of reference as well as the environment can support multiparty communicative goals. 

\section{Conclusion}
Amid the current pandemic there has been growing interest in how we can effectively work and collaborate remotely. Although VR seems like it should provide the perfect environment for such collaboration, we argue that currently VR is not adequately designed for such multiparty communicative contexts. Here we highlight how issues in social immersion, user roles and the design of effective spaces to incorporate share referents need to be addressed to ensure we improve the experience of VR users in such contexts. The outlining of these issues aims to act as a foundation for what to focus on when considering the design of multiparty VR spaces for conversation and collaboration.

\section{Acknowledgments}
This work was conducted with the financial support of the Science Foundation Ireland Centre for Research Training in Digitally-Enhanced Reality (D-REAL) under Grant No. 18/CRT/6224 and Science Foundation Ireland ADAPT Centre under Grant. No. 13/RC/2106.

%
\bibliographystyle{ACM-Reference-Format}
\bibliography{cui2020.bib}

\end{document}